# Vector Focus Wave Modes with Elliptic Cross-Section

Vitalis Vosylius and Sergej Orlov

*Center for Physical Sciences and Technology, Industrial laboratory for photonic technologies,*
*Sauletekio 3, LT-10257 Vilnius, Lithuania*
*E-mail: Sergejus,orlovas@ftmc.lt*

Nondiffracting pulsed beams are well studied nowadays and can be as short as a few femtoseconds. The nondiffracting pulsed beams not only resist diffraction but also propagate without changes due to the dispersion of a linear dispersive medium. A promising member of non-diffracting beam family is the so-called Mathieu beam which is a solution of Helmholtz wave equation in elliptical coordinate system. Zeroth order even Mathieu beams have unique asymmetric cross-section which makes these beams suitable for precise material processing. In this work we derive vectorial Mathieu beams using classical techniques and superpose them to create femtosecond pulsed beams. For these pulsed beams diffraction spreading and dispersive broadening is compensated by a given angular dispersion. Various intensity distributions, durations, angular dispersions and polarization states of different vector Mathieu focus wave modes are presented and discussed in detail.

## 1. Introduction

Since their introduction to the modern optics problems of localized wave propagation have attracted much attention. After the very first observation of the optical Bessel beam by Durnin et al. [1], number of publications in the topic of nondiffracting waves and their possible applications in various fields has exploded and is still growing. Free-space experimental realization of localized optical pulsed beams (Bessel X pulses, X-waves and focus wave modes) was achieved in practical setups [2,3]. Such pulses are polychromatic superposition of nondiffracting Bessel beams with focus wave modes being the most general case. Those pulsed beams propagate also without observable diffraction if the cone angles and frequencies of individual monochromatic Bessel beams that compose a pulsed beam areproperly related. Moreover, not only diffraction-free but also dispersion-free propagation of such optical pulses is possible in linear [2,4,5], as well as in nonlinear [6, 7] dispersive media. This phenomenon is usually observed when the cone angles and frequencies of the individual monochromatic Bessel beams that compose a pulsed beam are properly related.

Nondiffracting Bessel-like pulsed beams are well studied nowadays and can be as short as a few femtoseconds. The advantage of nondiffracting pulses is not only their nondiffracting property but also a dispersionless propagation inside a linear medium [4,5]. Best known Bessel pulsed beams are widely used in various fields like laser microprocessing [8].

Yet another member of a quite large family of nondiffracting beams is the elliptically shaped Mathieu beam which is a solution of Helmholtz wave equation in the corresponding coordinate system [9]. Zeroth order even Mathieu beams have a very promising elliptical cross-section, which makes these beams suitable for precise material processing [10] and lets us to call them "optical knives". Due to the elliptical coordinate system being able to provide us with an additional degree of freedom (eccentricity parameter) one can aim for the control of the width of such "optical knives". Moreover, besides the intensity control, another property of the electromagnetic field receives recently attention. Femtosecond vectorial and non-uniformly polarized pulsed beams are of significant interest due to their importance in material processing and growing applicability elsewhere [11,12].

In this work we start by deriving vector Mathieu beams using technique described in great detail in Refs. [13,14]. Afterwards we employ them to create a superposition which results in femtosecond pulsed beams with controllable polarization properties. For these nonhomogenously polarized pulsed beams diffraction spreading and dispersive broadening is compensated by a given angular dispersion as in the case of their Bessel-based counterparts.

## 2. Theory

We define elliptic cylinder coordinates by the transformation

$$x + iy = a \arccos(\xi + i\eta),$$
$$x = a \cosh \xi \cos \eta, \; \xi = [0, \infty),$$
$$y = a \sinh \xi \sin \eta, \; \eta = [0, 2\pi). \quad (1)$$

Here $a$ is ellipticity parameter, $\eta, \xi$ are transverse elliptic coordinates. In these coordinates the three-dimensional Helmholtz equation separates into a longitudinal and transverse parts as given in Ref. [9]

$$\psi_e(\mathbf{r}) = Je_m(\xi, q) ce_m(\eta, q) \exp(ik_z z), \quad (2)$$
$$\psi_o(\mathbf{r}) = Jo_m(\xi, q) se_m(\eta, q) \exp(ik_z z), \quad (3)$$

where, $Je_m$ is an even radial Mathieu function, $Jo_m$ - an odd radial Mathieu function, $ce_m$ is even angular Mathieu function and $se_m$ is an odd angular Mathieu function. Parameter $q = a^2 k_t^2 / 4$ and $k_t = k \sin \theta$ is a transverse wave vector component, $k_z = k \cos \theta$ is a longitudinal wave vector component of the wave vector $\mathbf{k}$, $\theta$ is the cone angle of the Mathieu beam and indices (e) and (o) correspond to the even and odd Mathieu beams of the order $m$, see [9]. Helical Mathieu beams are defined as [9]

$$\psi_h(\mathbf{r}) = \psi_o(\mathbf{r}) \pm i\psi_e(\mathbf{r}) \qquad (4)$$

In order to fully explain properties of nondiffracting pulses, full vectorial description must be used. Thus we vectorize scalar elliptic nondiffracting fields using [13,14]

$$\mathbf{M}(\mathbf{r},\omega,a) = \nabla \times [\mathbf{e}_z \psi_e(\mathbf{r},q)], \qquad (5)$$

$$k\mathbf{N}(\mathbf{r},\omega,a) = \nabla \times \mathbf{M}(\mathbf{r},\omega,a), \qquad (6)$$

where $\mathbf{e}_z = (0,0,1)$. An azimuthally polarized Mathieu beam is represented here by a vector field $\mathbf{M}$ and the radially – by a vector field $\mathbf{N}$.

$$M_x = \frac{a}{h^2}\big[Je_m(\xi,q)ce_m^{'}(\eta,q)\sinh(\xi)\cos(\eta)$$
$$+Je_m^{'}(\xi,q)ce_m(\eta,q)\cosh(\xi)\sin(\eta)\big]e^{i(k_z z - \omega t)},$$

$$M_y = \frac{a}{h^2}\big[Je_m(\xi,q)ce_m^{'}(\eta,q)\cosh(\xi)\sin(\eta)$$
$$-Je_m^{'}(\xi,q)ce_m(\eta,q)\sinh(\xi)\cos(\eta)\big]e^{i(k_z z - \omega t)},$$

$$M_z = 0.$$
$$(7)$$

$$N_x = \frac{ik_z a}{kh^2}\big[Je_m^{'}(\xi,q)ce_m(\eta,q)\sinh(\xi)\cos(\eta)$$
$$-Je_m(\xi,q)ce_m^{'}(\eta,q)\cosh(\xi)\sin(\eta)\big]e^{i(k_z z - \omega t)},$$

$$N_y = \frac{ik_z a}{kh^2}\big[Je_m^{'}(\xi,q)ce_m(\eta,q)\cosh(\xi)\sin(\eta)$$
$$+Je_m(\xi,q)ce_m^{'}(\eta,q)\sinh(\xi)\cos(\eta)\big]e^{i(k_z z - \omega t)},$$

$$N_z = \frac{4q}{a^2 k}Je_m(\xi,q)ce_m(\eta,q)e^{i(k_z z - \omega t)}.$$
$$(8)$$

Here $\omega$ frequency of the electromagnetic field. Nondiffracting vector Mathieu pulsed beams are introduced here as a polychromatic superposition of vector Mathieu elliptic beams. Thus, vector elliptic (or Mathieu) focus wave modes are described as

$$\mathbf{E}(\mathbf{r},t) = \int_0^\infty S(\omega)\mathbf{E}_0(\mathbf{r};\omega)\exp(-i\omega t)d\omega \qquad (9)$$

where $\mathbf{E}_0(\mathbf{r},\omega)$ is a complex amplitude of vector monochromatic beam (see, Eqs. (5,6)) with frequency $\omega$.

An integral (9) is evaluated in general only numerically. This is due to the fact that focus wave modes are the most general case of dispersionless and diffractionless pulses and those pulses can have either a positive or a negative angular dispersion $d\theta/d\omega$ of the cone angle θ. In the ideal case FWMs propagate in the air or inside a dispersive medium and its group velocity remains constant over the whole spectral range $cdk/dk_z = V$. Thus, the longitudinal projection $k_z$ of the wave vector must satisfy relation [4,5]

$$k_z(\omega) = \frac{\omega}{c}n(\omega)\cos\theta = \frac{\omega}{V} + \gamma \qquad (8)$$

where $V$ is a group velocity of the pulse and $\gamma$ – separation constant, $c$ – velocity of light in vacuum, $n(\omega)$ –

refractive index of a material. From (8) the angular dispersion of the pulse can be explicitly expressed:

$$\theta(\omega) = \arccos\left(\frac{c}{Vn(\omega)} + \frac{\gamma c}{\omega n(\omega)}\right) \qquad (9)$$

In the case, when $\gamma = 0$, there is no angular dispersion present, and the group velocity of the pulse coincides with its phase velocity and is always superluminal

$$V = \upsilon_f = c/\cos\theta \qquad (10)$$

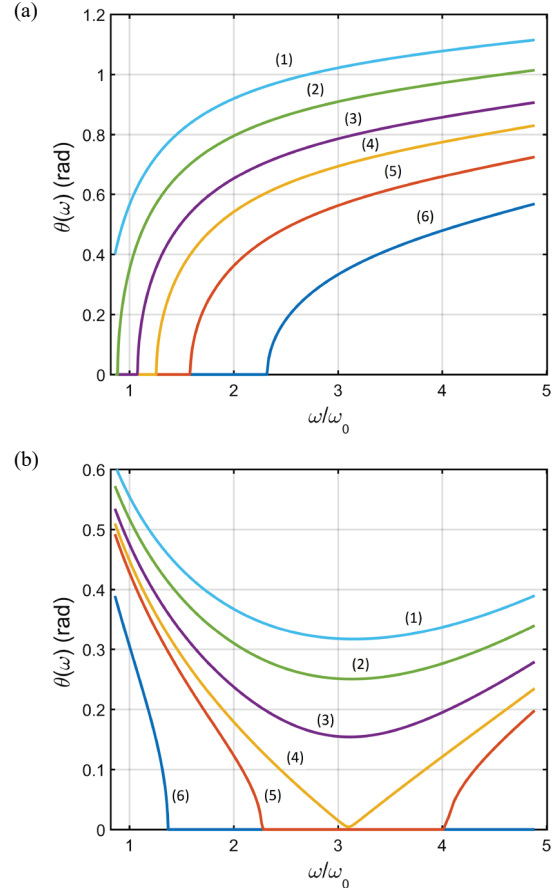

**Fig. 1** Dispersion curves of diffraction free parabolic pulses in Sapphire. Parameter (a) $\gamma$=5 μm⁻¹, V/c=1.5 (1), V/c=1.2 (2), V/c=1 (3), V/c=0.9 (4), V/c=0.8 (5), V/c=0.7 (6) (b) $\gamma$=5 μm⁻¹, V/c=0.55 (1), V/c=0.54 (2), V/c=0.53 (3), V/c= 0.52407 (4), V/c=0.52 (5), V/c = 0.5 (6), $\omega_0$=1.83 ×10¹⁵s⁻¹.

When $\gamma \neq 0$, the phase velocity of the pulse can be expressed as

$$\upsilon_f = \frac{\omega V}{\omega + \gamma V} \qquad (11)$$

Moreover, for the sake of brevity we restrict ourselves in this report with the case of a radially polarized FWMs, described by the Eq. (6, 8, 9). We also integrate in Eq. (9) over a rectangular spectral amplitude with the center located at the carrier frequency $\omega_c$ and with spectral width $\Delta\omega$.

$$\mathbf{E}(\mathbf{r},t) = \int_{\omega_0 - \Delta\omega/2}^{\omega_0 + \Delta\omega/2} S(\omega)\mathbf{N}(\mathbf{r},\omega,a)\exp(-i\omega t)d\omega \qquad (12)$$

The case of the azimuthally polarized Mathieu-type FWMs will be presented elsewhere.

## 3. Numerical results

The typical angular dispersion curves in sapphire for different values of $V/c$ for a single value of $\gamma$ are shown in Fig. 1. The frequency $\omega$ is normalized to frequency $\omega_0$, which is the carrier wavelength of laser system routinely used in the lab. We fix the parameter $\gamma$ as it ensures a rather rich choice of different angular dispersion curves for different values of $V/c$, see Fig. 1.

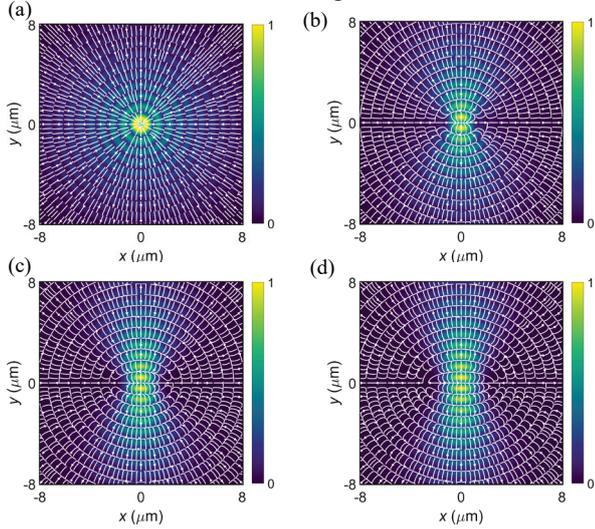

**Fig. 2** Intensity distributions of the radially polarized Mathieu FWM inside sapphire in the transverse plane for q = 0 (a), q = 3 (b), q = 10 (c) and q = 20 (d). Parameters are $\gamma$= 5 $\mu m^{-1}$, $V/c$ = 1.2, $\omega_c$ = 1.83×10¹⁵ s⁻¹, τ = 50 fs. White lines represent the orientation of the electric field.

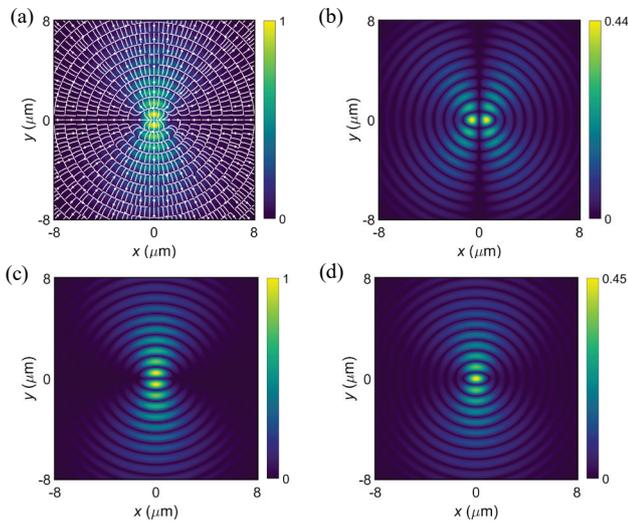

**Fig. 3** Intensity distributions of the radially polarized Mathieu FWM and its linearly polarized components inside sapphire in the transverse plane for $q$ = 3, $\gamma$= 5 $\mu m^{-1}$, $V/c$ = 1.2, $\omega_c$ = 1.83×10¹⁵ s⁻¹ τ = 50 fs.

Before we proceed with discussion on our numerical results we note, that sapphire is a birefringent uniaxial material. The difference between refractive indices of ordinary and extraordinary beams in our region of interest is $\Delta n = 10^{-5}$. We choose the direction of the crystal axis that coincides with the axis of propagation z. In this case the azimuthal polarization is the ordinary beam and the radial polarization is the extraordinary beam.

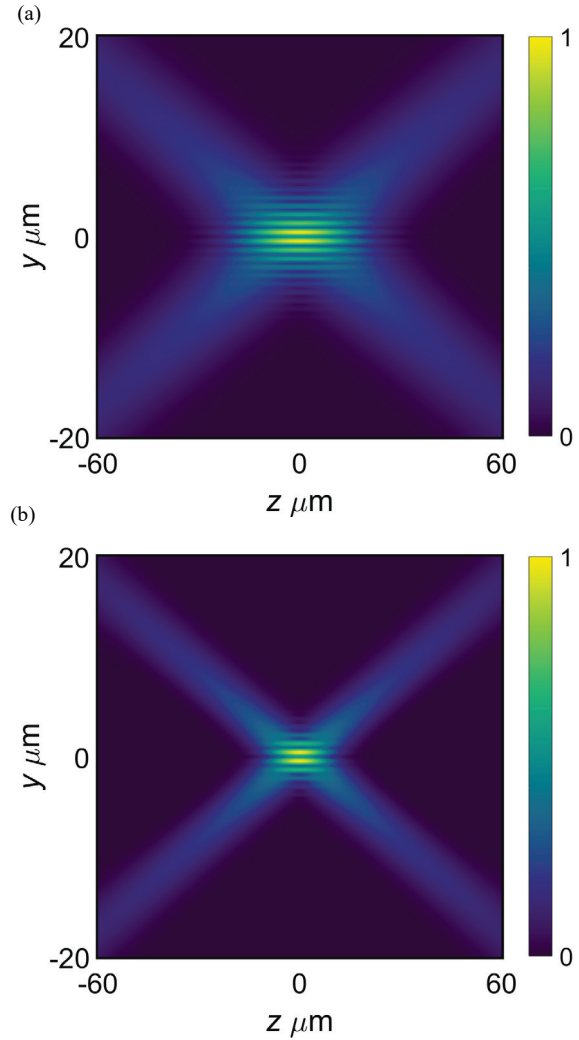

**Fig. 4** Intensity distributions of the radially polarized Mathieu FWM inside sapphire in the longitudinal plane for. $q$ = 3, $\gamma$ = 5 $\mu m^{-1}$, $V/c$ = 1.2, $\omega_c$ = 1.83×10¹⁵ s⁻¹. Durations are τ = 50 fs (a) and τ = 25 fs (b).

For the sake of brevity we restrict ourselves with the case of a radially polarized FWMs, described by the Eq. (8, 12). We note, that for angles and wavelengths depicted in Fig. 1 the refractive index of the extraordinary beam is equal to the refractive index of the ordinary beam and the birefringence of the sapphire can be neglected. We also integrate in Eq. (12) over a rectangular spectral amplitude $S(\omega) = \Pi(2(\omega - \omega_c)/\Delta\omega)$ (here $\Pi$ is a boxcar function) with the center located at the carrier frequency $\omega_c$ and with spectral width $\Delta\omega$.

We start with cases of increasing ellipticity parameter $q$ (from $q$=0, Bessel beam, to $q$=20, two counter propagating Gaussian beams) and choose the fourth dispersion curve (4) in the Fig. 1 (b) and plot the electric field distributions in the transverse plane, see Fig 2.

First, we note that the polarization structure resembles that of a radially polarized beam and does not change a lot with when we increase the ellipticity of the beam, though for extreme values of the parameter $q$ the polarization pat-tern becomes elliptically distorted in the direction perpendicular to the intensity axis, see Fig 2.

This effect can be understood while analyzing individual components of the radially polarized elliptic FWM, see Fig. 3. First of all, due to the tight focusing and components with large angles to the propagation axis, we observe an appearance of a relatively strong $z$ component. Two transverse components ($x$ and $y$) have different shapes and also slightly different maximal values, see Fig. 3. However maximal values of the $y$ component are situated not on $x$-axis but on the cross which has angle larger than 45 degrees. Therefore, we observe some distortions in the polarization pattern of the radially polarized beam, which is known to be rather symmetric for Gaussian modes.

As a next step, we look at the two cases with different durations, see Fig. 4. First, what we note here, is that the elliptic FWM of a smaller duration can be actually larger in the transverse plane than the optical field with larger duration. At first this might to look contraintuitive, because we are forgetting the fact, that those optical fields have a distinct angular dispersion. This dispersion is described by the curve (4) in the Fig. 1 (b). When the bandwidth of the signal becomes larger, i.e. the case of smaller duration, we have to include cone angles of plane wave components which are smaller than those in the case of a larger duration. Because of that an effective defocusing of the spatio-temporal profile happens and the beamwidth might increase, exactly as seen in Fig. 4.

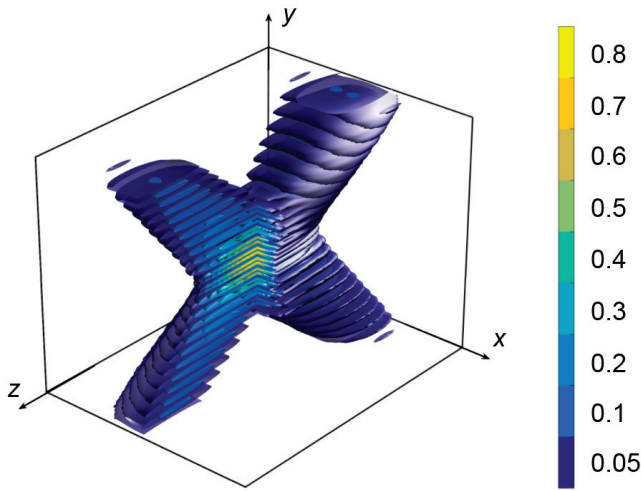

**Fig. 5** A three dimensional depiction of an intensity isosurface of the radially polarized elliptic FWM inside sapphire, $q = 3$, $\gamma = 5$ μm$^{-1}$, $V/c = 1.2$, $\omega_c = 1.83 \times 10^{15}$ s$^{-1}$ and $\tau = 50$ fs.

Additionally, we demonstrate here a three dimensional distribution of the intensity profile, see Fig 5. Obviously, as long as the ellipticity is increasing, we observe an asymmetrical X letter shape for smaller intensities due to the nondiffracting nature and a Mathieu like modulation for larger intensities.

An interesting comparison here would be to compare elliptic beams with their Bessel counterparts, which are known to change their X-shape into an O-shape, when the dispersion curve has a distinct minima, see Ref. [5] for more information.

We stick now to the case of a small ellipticity from Fig. 2 and move the carrier frequency onto the point on the dis-persion curve (4) from the Fig. 1, where the angle is smallest. Our numerical simulation results are depicted in Fig. 6.

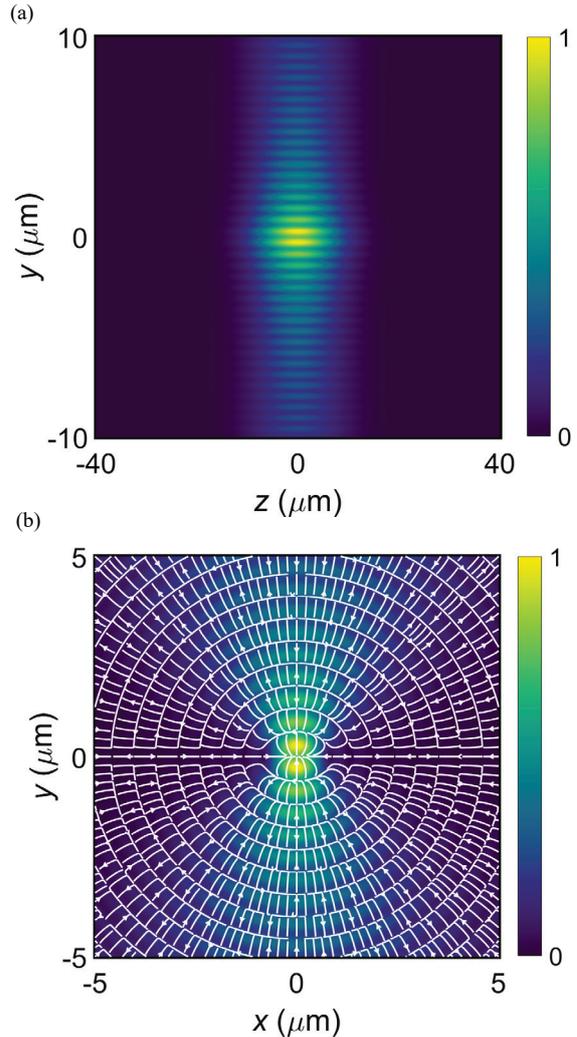

**Fig. 6** Intensity distributions of the radially polarized Mathieu FWM inside sapphire in the longitudinal plane (y,z) (left) and in the transverse plane (x,y) (right) for q = 3, $\gamma$= -2 μm$^{-1}$, $V/c$ = 0.55, $\omega_c$ = 1.83×10$^{15}$ s$^{-1}$ and τ = 50 fs.

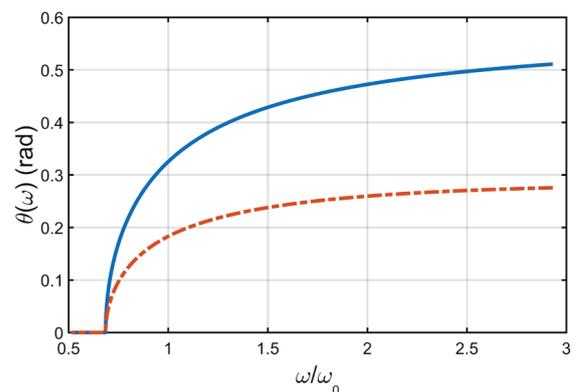

**Fig. 7** Angular dispersion curves of a FWM in the air (1) and sapphire (2). Parameters are γ= 0.7 μm$^{-1}$, $V/c$ = 1.2, $\omega_0$ = 1.83×10$^{15}$ s$^{-1}$.

First, we note that the transverse dimensions have increased here due to the smaller angles of the individual plane wave components, i.e. smaller spatial frequencies. However, the shape of the Mathieu-type FWM has experienced some changes. Now we don't see the distinct X-letter shape in the region of smaller intensities and the intensity distribution in the longitudinal plane is reminiscent of the letter O, like in the case of classical Bessel-type focus wave modes.

Lastly, we analyze here a propagation of radially polarized elliptic FWMs from air to sapphire. Propagation of pulses without group velocity dispersion is ensured by a strictly defined angular dispersion, see Eq. (8,9). We choose here a superluminal group velocity $V = 1.2c$ of the pulse in air, the parameter $\gamma = 0.3 \; \mu m^{\wedge}(-1)$ is the same as previously. Obviously, upon the transmittance through the interface the angular dispersion inside the glass is affected by the Snell's law and by the material dispersion, see Fig. 7.

Intensity distributions of a radially polarized Mathieu type FWM entering the sapphire from the air are depicted in the Fig. 8. Pulses are depicted at different time moments before and after they enter the sapphire. After the FWMs cross the surface, one can clearly distinguish between reflected and transmitted pulses, see Fig 8. As the radially polarized FWM pulse passes the interface, its intensity is continuous on the interface. An interesting effect is observed as the radially polarized pulse is defocused inside the sapphire, so the energy density at the center of the pulsed beam decreases. This happens due to the decrease in the angular spectrum of the FWH, which happens because the angular dispersion of the pulse is affected, see Fig. 7. Transmitted FWMs undergo group velocity dispersion that causes them to spread. After $t = 200$ fs in sapphire, durations of radially polarized pulsed beams would increase. Further propagation causes pulses to split into two intensity peaks as higher order material dispersion kicks in.

In order to overcome the dispersion spread of the FWM pulses inside the sapphire, we have to introduce an additional dispersion to the pulses in the air before they enter the glass. This can be achieved by changing the angular dispersion in Eq. (7) so, that the FWM pulse retains its nondiffracting and nondispersing properties after it enters the glass. An example of this dispersion compensation is presented in the Fig. 7. In this case, after the pulses enter from the air into the sapphire they propagate without any group velocity dispersion.

## 4. Conclusion and discussion

In conclusion, we have studied the propagation of radially polarized elliptic (Mathieu) FWMs inside the sapphire. We have revealed both the influence of the ellipticity parameter and of the carrier frequency on the shape of the Mathieu-type radially polarized focus wave mode. Lastly we have covered problems arising when a parabolic FWM enters the sapphire from the air through a planar interface. A dispersive pulse spreading is observed due to material dispersion and due to the change in the angular dispersion.

Nonhomogeneous beam profiles and polarizations like azimuthal and radial are becoming important in areas of laser microprocessing, as the number of reports on their efficiency for material processing applications is increasing [10]. For example, ablation rate of azimuthally or radially polarized pulses are measured to be higher in comparison to frequently used linearly or circularly polarized light [11].

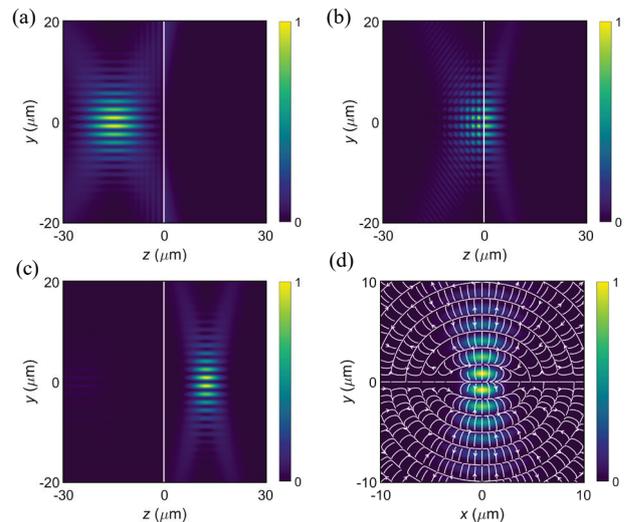

**Fig. 8** Intensity distributions of a radially polarized FWM at different moments in time: (a) $t = 15$ fs before entering medium (b) at the center of the interface between the air and sapphire, (c) $t = 25$ fs after the entering the medium. Illustration (d) depicts situation in the transverse plane (x,y) just after FWM enters the medium. Parameters are $\gamma = 0.7 \; \mu m^{-1}$, $V/c = 1.2$, $\omega_c = 1.83 \times 10^{15}$ s$^{-1}$

introduced by the interface. We have introduced also a strategy to account for effects due to material dispersion

Moreover, azimuthally polarized pulses are perspective for achieving better ablation quality when drilling high aspect ratio holes [12]. Elliptic cross-section of a Bessel-type beam also is helpful in laser assisted cracking of glasses [15].

Moreover, in perspective, higher topological charges may introduce novel effects due to their relation to the angular momentum of light [16], thus focus wave modes with higher orders are of interest also. They might also lead to novel structured patters, important in various laser related micro/nanoprocessing application as it was the case for superpositions of Bessel beams [17].

Further investigations on applications of parabolic azimuthally and radially focused wave modes with femtosecond durations will be presented elsewhere.

## Acknowledgment

This research is/was funded by the European Social Fund according to the activity 'Improvement of researchers' qualification by implementing world-class R&D projects' of Measure No. 09.3.3-LMT-K-712.